\documentclass{moriond}

\def\mco{\multicolumn}

\def\be{\begin{equation}}
\def\ee{\end{equation}}
\def\bea{\begin{eqnarray}}
\def\eea{\end{eqnarray}}

\usepackage{wrapfig}
\usepackage{placeins}
\usepackage[hang,flushmargin]{footmisc}

\newcommand\blfootnote[1]{%
  \begingroup
  \renewcommand\thefootnote{}\footnote{#1}%
  \addtocounter{footnote}{-1}%
  \endgroup
}

\newcommand{\Photo}{}

\begin{document}
\vspace*{4cm}
\title{AN EXPERIMENT EXPLORING GRAVITATIONAL EFFECTS \\ON CP VIOLATION}

\author{ G.M.~PIACENTINO\,$^{*,1,2,3}$,
         A.~PALLADINO\,$^{\dag,4}$,
         R.N.~PILATO$^{5,6}$,
         G.~VENANZONI$^{6}$,
         L.~CONTI$^{1,2}$,
         G.~DI~SCIASCIO$^{2}$,
         R.~DI~STEFANO$^{7}$,
         N.~FRATIANNI$^{2,8}$,
         A.~GIOIOSA$^{6,8}$,
         D.~HAJDUKOVIC$^{9}$,
         F.~IGNATOV$^{10}$,
         F.~MARIGNETTI$^{8}$,
         V.~TESTA$^{3}$
       }

\address{$^{1}$Uninettuno University, Rome, Italy,\\
$^{2}$INFN, Sezione di Roma Tor Vergata, Rome, Italy,\\
$^{3}$INAF, Osservatorio Astronomico di Roma, Monteporzio Catone, Italy,\\
$^{4}$Boston University, Boston, USA,\\
$^{5}$Dipartimento di Fisica, Università di Pisa, Pisa, Italy,\\
$^{6}$INFN, Sezione di Pisa, Pisa, Italy,\\
$^{7}$INFN, Sezione di Napoli, Naples, Italy,\\
$^{8}$Università degli Studi del Molise, Campobasso, Italy,\\
$^{9}$INFI, Cetinje, Montenegro,\\
$^{10}$BINP, Novosibirsk, Russia}

\maketitle\abstracts{
We suggest a new experiment sensitive to a possible difference between the amount of CP violation as measured on the surface of the Earth and in a lower gravity environment. Our proposed experiment is model independent and could yield a $5\sigma$ measurement within tens of days, indicating a dependence of the level of CP violation in the neutral kaon system on the local gravitational potential.
}

\section{Introduction}
Einstein’s weak equivalence principle (WEP) appears to hold\,\cite{huber-2000,huber-2001} for all types of matter with an accuracy of one part in $10^{11}$, however, several authors suggest that it does not hold for antimatter\,\cite{chardin-1993,villata-2011}.
Instead of a direct measurement of the gravitational mass of antimatter, we are concerned with the possibility, suggested by Chardin\,\cite{chardin-1993} and implicit in the work of Good\,\cite{good-1961}, that a repulsive interaction between matter and antimatter could be a (beyond Standard Model) source of CP violation, at least in the neutral kaon system. Our proposed experiment takes advantage of the large difference between the gravitational field on the surface of the Earth and on the Moon, in a low-Earth orbit (LEO), or a Lagrange-point orbit (LPO).
Since the branching ratio of the CP-violating decay, $R = \frac{\Gamma(\mathrm{K}_{\mathrm{L}}\to \pi^+\pi^-)}{\Gamma(\mathrm{K}_{\mathrm{L}}\to \pi^+\pi^-\pi^0)}$,
is quadratic in $\varepsilon$, we expect a very large difference in the value of $R$ when measured in space as compared to the measurement on Earth.
This measurement of CP violation as a function of the local gravitational curvature of space-time could provide two very significant scientific results:
(a) an indirect probe of the gravitational interaction of antimatter; and
(b) an implication of a new source of CP violation other than the weak interaction.\blfootnote{
              Contribution to the 2021 Gravitation session of the 55$^\mathrm{th}$ Rencontres de Moriond\\
              Corresponding authors:
              $^{*}$giovannimaria.piacentino@uninettunouniversity.net,
              $^{\dag}$palladin@bu.edu
              }

\section{Physics reach / Outcome}
The induced CP violation within the neutral kaon system is described, for predictive purposes, using semi-classical approximation in which the gravitational field is classical and the transitions of the quarks are quantum phenomena. Our experiment could therefore be the first measurement to explore quantum effects of gravity and also the first to provide quantitative indications of a possible quantization of the gravitational interaction, even in the case of a null result.

The number of particle physics experiments performed since the fifth decade of the twentieth century is enormous, however none of them have considered possible gravitational systematic effects, despite having been carried out on Earth and therefore exposed to a relatively intense gravitational field. The experiment we propose would be the first to explore this possible systematic effect which may not necessarily be negligible since each particle interacts with the whole Earth.

Finally, this hypothesis may overcome many of the open problems in astrophysics and cosmology.
The Cosmic Microwave Background is neither anisotropic nor inhomogeneous enough to be compatible with the Big Bang model without the introduction of an unknown interaction driving inflation. The globally repulsive gravitational interaction inside a newborn universe made of equal contribution of matter and antimatter could have driven inflation and a strong CP violation due to the early intense gravitational field could have caused the matter/antimatter asymmetry, according to Sakharov's hypothesis\,\cite{sakharov-1967}. Speculative alternative cosmological models based on gravitational repulsion between matter and antimatter are already being developed\,\cite{hajdukovic-2020}.

\section{Scientific background}
\vspace{-0.5cm}
\begin{table}[!h]
\begin{tabular}{lc}
\begin{minipage}{0.58\textwidth}
Matter is gravitationally self-attractive:
\vspace{0.3cm}
\end{minipage} &
\begin{minipage}{0.38\textwidth}
\centering $\frac{d^2x^\lambda}{d\tau^2} = - \frac{dx^\mu}{d\tau}\frac{dx^\nu}{d\tau} \Gamma^{\lambda}_{\mu\nu}$
\vspace{0.3cm}
\end{minipage} \\
\begin{minipage}{0.58\textwidth}
\vspace{0.6cm}
According to general relativity, after applying the\\
CPT transformation ($dx^{\mu} \to -dx^{\mu}$ and $\Gamma^{\lambda}_{\mu\nu} \to -\Gamma^{\lambda}_{\mu\nu}$),\\
antimatter should also be gravitationally self-attractive:
\vspace{0.5cm}
\end{minipage} &
\begin{minipage}{0.38\textwidth}
\vspace{0.8cm}
\centering $-\frac{d^2x^\lambda}{d\tau^2} = - \left( -\frac{dx^\mu}{d\tau} \right) \left( -\frac{dx^\nu}{d\tau} \right) \left( -\Gamma^{\lambda}_{\mu\nu} \right)$
\vspace{0.3cm}
\end{minipage} \\
\begin{minipage}{0.58\textwidth}
There is no experimental conclusion yet whether there\\
exists an attraction or repulsion between matter and\\
antimatter. General relatively predicts a \emph{repulsion} \\
when CPT is applied to one object but not the other\,\cite{villata-2011}:
\end{minipage} &
\begin{minipage}{0.38\textwidth}
\vspace{1.3cm}
\centering $-\frac{d^2x^\lambda}{d\tau^2} = - \left( -\frac{dx^\mu}{d\tau} \right) \left( -\frac{dx^\nu}{d\tau} \right) \Gamma^{\lambda}_{\mu\nu}$
\vspace{0.05cm}
\end{minipage}\\
\end{tabular}
\end{table}
\FloatBarrier

\section{Motivation}

\begin{wrapfigure}{r}{0.35\textwidth}
  \begin{center}
    \vspace{-1.25cm}
    \includegraphics[width=0.33\textwidth]{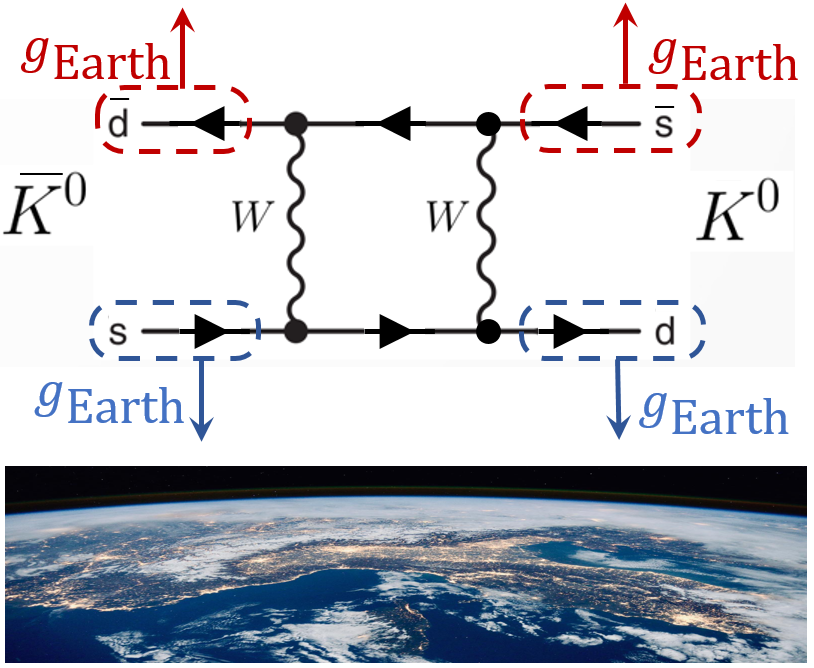}
  \end{center}
  \caption{Within the neutral kaon system, the matter components could be attracted to Earth while the antimatter components are repelled.}
  \label{fig:feynman}
\end{wrapfigure}
A gravitationally-induced separation of the matter and antimatter components of the neutral kaon (Figure~\ref{fig:feynman}) would be proportional to the mixing time, $\Delta\tau = \frac{\pi\hbar}{\Delta m_\mathrm{K}c^2} \simeq 5.9 \times 10^{-10}$~s, resulting in an induced separation of $\Delta\xi \sim g\tau^{2}$. This separation would cause a regeneration of the $\mathrm{K}_\mathrm{S}$ component, providing a source for the observed CP-violating $2\pi$ decay mode.

The amount of CP violation induced by this phenomenon would be
$\chi = \frac{\Delta\xi}{L_{\mathrm{Compton}}} \sim O(1) \times g \frac{\hbar m_{\mathrm{K}} c}{(\Delta m_\mathrm{K} c^2)^2} \sim O(1) \times0.88\times10^{-3}$. Interestingly, this happens to be the same order of magnitude as the level of CP violation observed on the surface of Earth, $\varepsilon \simeq 2.2 \times 10^{-3}$.
\FloatBarrier

\section{Detector}

\begin{wrapfigure}{rht!}{0.40\textwidth}
  \begin{center}
    \vspace{-1.25cm}
    \includegraphics[width=0.38\textwidth]{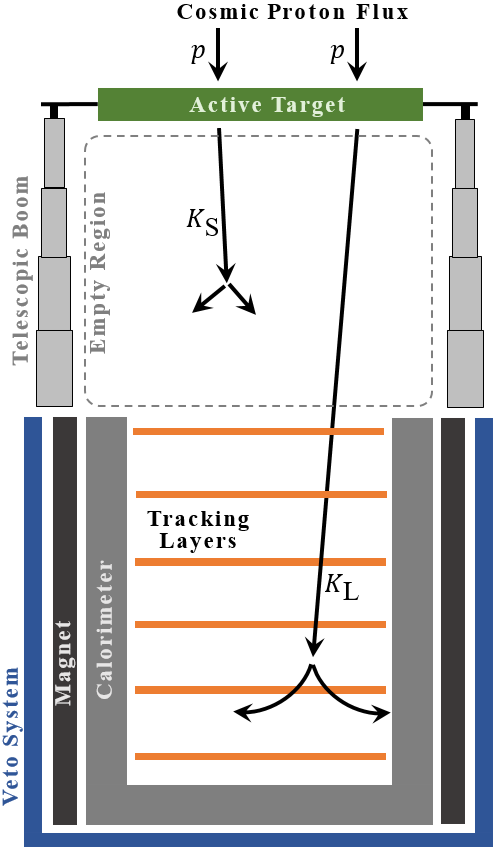}
  \end{center}
  \caption{Detector consisting of an active target, empty region, and a tracking region surrounded by an electromagnetic calorimeter, a magnet, and a veto system.}
  \label{fig:detector}
\end{wrapfigure}
Our proposed measurement can be achieved using an experimental apparatus (Figure~\ref{fig:detector}) placed in a low-gravity environment.
The detector consists of a cylindrical \emph{active target}, $\sim$20~cm deep, consisting of alternating layers of scintillating material and PbWO$_4$.
Our simulations\,\cite{piacentino-2016} show that cosmic protons incident upon a PbWO$_4$ target would produce a reasonable rate of $\mathrm{K}_\mathrm{L}$. The PbWO$_4$ can be substituted with lunar regolith, if the experiment is deployed on the surface of the Moon to reduce transportation weight.
Telescopic booms could extend the active target away from the tracking region, providing an \emph{empty region} in which the $\mathrm{K}_\mathrm{S}$ are allowed to decay, suppressing the $\mathrm{K}_\mathrm{S} \to 2\mathrm{\pi}$ background. The active target could be brought closer to the tracking region for systematics studies and for other physics experiments, extending the operational life of the detector in space.
A cylindrical \emph{tracking volume} (1~m diameter, 1~m deep) consists of silicon tracking layers.
A cylindrical magnet would allow identification of charged particles and help distinguish between pions and muons.
An electromagnetic calorimeter could provide a precise energy resolution to further distinguish between pions and muons.
A \emph{veto system} consisting of scintillating detectors surrounding the whole tracking volume would further suppress cosmic background contamination.

\FloatBarrier
\section{Simulation}

We performed a Geant4 simulation using the angular and energy spectrum of the incident cosmic protons in space, as estimated by AMS-02\,\cite{ams-detector} and PAMELA\,\cite{pamela-2017} (for LEO) and CRaTER\,\cite{ackerman-2016} (for the Moon's surface).
The $\mathrm{K}_{\mathrm{S}}$ background contamination can be significantly reduced by selecting only $\mathrm{K}_{\mathrm{S,L}}$ that decay with low forward momentum (e.g., $p_z<1~GeV$) with minimal loss in the number of signal $\mathrm{K}_{\mathrm{L}}$ decays, as described in~\cite{piacentino-2016,piacentino-2017,piacentino-2019}. An additional key background will come from any confusion between $\mathrm{K}_{\mathrm{L}} \to \pi\mu\nu$ and $\mathrm{K}_{\mathrm{L}} \to \pi\pi$ decays. We are confidant that
modern detector technology will allow us to sufficiently distinguish between pions and muons causing only a modest increase in the required measurement time. Assuming perfect particle identification, our {\textsc Geant4} simulations indicate that it would take $\mathcal{O}( \mathrm{days} )$ ($\mathcal{O}\left(\mathrm{tens\,of\,days}\right)$) to record sufficient $\mathrm{K}_{\mathrm{L}}$ decays to provide 3$\sigma$ (5$\sigma$) measurements of $R$.
\begin{table}[h!]
\caption[]{Requirements for 3$\sigma$ and 5$\sigma$ measurements of $R$ in low gravity environments assuming either a linear dependence of $\varepsilon$ on $g$, or that $\varepsilon$ is independent of $g$.}
\label{tab:results}
\vspace{-0.25cm}
\begin{center}
\begin{tabular}{|l|c|c|c|c|}
\hline
\mco{1}{|c|}{Measurement} &
\mco{2}{|c|}{$N\left(\mathrm{K}_\mathrm{L} \, \mathrm{decays}\right)$} &
\mco{2}{|c|}{$T_\mathrm{min}$ to collect sufficient $\mathrm{K}_\mathrm{L}$ decays}
\\
{} & $3\sigma$ & $5\sigma$ & $3\sigma$ & $5\sigma$
\\ \hline
\begin{minipage}{2.1in}
\vspace{0.05cm}
$R$ on the surface of the Moon,\\if $\varepsilon \propto g$
\vspace{0.05cm}
\end{minipage} &
3.3 $\times$ $10^{5}$ &
9.1 $\times$ $10^{5}$  &
\begin{minipage}{1.1in}
\centering 158 days
\end{minipage} &
439 days
\\ \hline
\begin{minipage}{2.1in}
\vspace{0.05cm}
$R$ in a Low Earth Orbit,\\if $\varepsilon \propto g$
\vspace{0.05cm}
\end{minipage} &
1.1  $\times$ $10^{4}$ &
3.1  $\times$ $10^{4}$ &
6 days &
15 days
\\ \hline
\begin{minipage}{2.1in}
\vspace{0.05cm}
$R$ in either LEO or on the Moon,\\if $\varepsilon$ is independent of $g$
\vspace{0.05cm}
\end{minipage} &
9.0  $\times$ $10^{3}$ &
2.5  $\times$ $10^{4}$ &
5 days &
12 days
\\ \hline
\end{tabular}
\end{center}
\end{table}

\FloatBarrier
\vspace{-1.5cm}
\section{Comparisons to other experiments}
\vspace{-0.25cm}
The first attempt to measure the direction of gravitational pull (or push) on antimatter was made by Wittenborn and Fairbank\,\cite{wittenborn-1967} measuring the free fall of e$^-$ and e$^+$ in 1967.
In 1986, a Los Alamos-led team proposed\,\cite{jarmie-1986} to measure the gravitational force on antiprotons at LEAR (CERN). In 1997, Phillips\,\cite{phillips-1996} suggested an interferometric experiment (never done) at Fermilab. The experimental constraints\,\cite{alpha-2013} on the gravitational push or pull between matter and antimatter is $-65 g_\mathrm{Earth} < g_{m,\bar{m}} <  110 g_\mathrm{Earth}$. The three ongoing experiments at CERN (AEGIS, ALPHA-g, and GBAR) involve antihydrogen. A recent proposal\,\cite{mage-2018,kaplan-2020} was submitted for the first experiment involving leptons.
Our approach benefits from: low systematics; not sensitive to binding energy; no need to create, confine, or transport anti-atoms; and no new technology is required (standard particle physics detectors suffice).
The main difficulty is placing/operating the detector in space.

\vspace{-0.35cm}
\section{Conclusions}
\vspace{-0.25cm}
By placing a detector in a low-Earth orbit, Lagrange-point orbit, or on the surface of the Moon, we could perform a direct measurement of the ratio of the number of $\mathrm{K}_\mathrm{L}$ decaying to two pions to those decaying to three pions in a low-gravity environment. We estimate that it will take $\mathcal{O}( \mathrm{days} )$ ($\mathcal{O}\left(\mathrm{tens\,of\,days}\right)$) to record sufficient $\mathrm{K}_{\mathrm{L}}$ decays for a 3$\sigma$ (5$\sigma$) measurement of $R$. Any difference between the amount of CP~violation in a low gravity environment with respect to the CP~violation on the surface of Earth could be an indication of a quantum gravitational effect.


\vspace{-0.34cm}
\section*{References}
\vspace{-0.2cm}
\begin{thebibliography}{99}

\bibitem{huber-2000}
F.M. Huber, R.A. Lewis, E.W. Messerschmid, G.A. Smith, \emph{Precision tests of Einstein's Weak Equivalence Principle for antimatter} Adv. in Space Res., {\bf 25}, 6, (2000) 1245–1249.

\bibitem{huber-2001}
F.M. Huber, E.W. Messerschmid, G.A. Smith, \emph{The WEAX experiment}, Class. Quantum Grav. {\bf 18} (2001) 2457–2466.

\bibitem{chardin-1993}
G. Chardin, \emph{CP violation and antigravity (revisited)},
Nuclear Physics A {\bf 558} (1993) 477c.

\bibitem{villata-2011}
M.~Villata, \emph{CPT symmetry and antimatter gravity in general relativity}, EPL {\bf 94}, 2  (2011).

\bibitem{sakharov-1967}
A.D. Sakharov. \emph{Violation of CP invariance, C asymmetry, and baryon asymmetry of the universe}. Journal of Experimental and Theoretical Physics Letters. {\bf 5} (1967) 24.

\bibitem{hajdukovic-2020}
D.S. Hajdukovic. \emph{Antimatter gravity and the Universe}, Mod. Phys. Lett. A {\bf 35}, (2020).

\bibitem{good-1961}
M. L. Good, \emph{K$^0_2$ and the Equivalence Principle},
Phys.Rev. {\bf 121} (1961) 311--313.

\bibitem{piacentino-2016}
G.M.~Piacentino, A.~Palladino, G.~Venanzoni, \emph{Measuring gravitational effects on antimatter in space}, Physics of the Dark Universe, {\bf 13}, (2016) 162--165. 

\bibitem{piacentino-2017}
G.M.~Piacentino, A.~Gioiosa, A.~Palladino, G.~Venanzoni, \emph{Measuring gravitational effects on antimatter in space}, ADMPP'16, EPJ Web of Conferences, {\bf 142}, (2017) 01023.

\bibitem{piacentino-2019}
G.M.~Piacentino, A.~Gioiosa, A.~Palladino, V.~Testa, G.~Venanzoni. \emph{Probing antigravitational effects through CP violation on the Moon},  (2019) {\verb arXiv: 1907.06866 [astro-ph.HE]}


\bibitem{ackerman-2016}
M.~Ackerman et al., \emph{Measurement of the high-energy gamma-ray emission from the Moon with the Fermi Large Area Telescope}, Phys. Rev. D {\bf 93} 8, 082001 (2016).

\bibitem{ams-detector}
M. Aguilar et al., Phys. Rev. Lett {\bf 110} (2013) 141102.

\bibitem{pamela-2017}
O.~Adriani, et al., \emph{Ten Years of PAMELA in Space}, Riv.\,del Nuovo Cim., {\bf 10}, (2017).

\bibitem{wittenborn-1967}
F.C. Witteborn, W.M. Fairbank, \emph{Experimental Comparison of the Gravitational Force on Freely Falling Electrons and Metallic Electrons}, Phys. Rev. Lett., {\bf 19}, (1967) 1049.

\bibitem{jarmie-1986}
N. Jarmie, \emph{A Measurement of the Gravitational Acceleration of the Antiproton: An Experimental Overview}, 8$^\mathrm{th}$ Conf. on the Appl. of Accelerators in Res. and Ind., (1986).

\bibitem{phillips-1996}
T.J. Phillips, \emph{Measuring the gravitational acceleration of antimatter with an antihydrogen interferometer}. Hyperfine Interactions {\bf 100}, (1996) 163–-172.

\bibitem{alpha-2013}
The ALPHA Collaboration., \emph{Description and first application of a new technique to measure the gravitational mass of antihydrogen}. Nature Communications {\bf 4}, (2013) 1785. 

\bibitem{mage-2018}
MAGE Collaboration, \emph{Studying Antimatter Gravity with Muonium}, Atoms, {\bf 6}, 2, (2018).

\bibitem{kaplan-2020}
D.M.~Kaplan, et al. \emph{Letter of Interest for a Muonium Gravity Experiment at Fermilab}, July 29, 2020.

\end{thebibliography}
\end{document}



